\documentclass{ws-procs9x6}

\begin{document}

\title{Search for Excited Neutrinos at HERA }

\author{C.~Diaconu\\ on  behalf of the H1 Collaboration}

\address{Centre de Physique des Particules de Marseille, 13288\\
E-mail: diaconu@cppm.in2p3.fr}

\maketitle

\abstracts{
A search for excited neutrinos produced in electron--proton collisions is performed using a data sample corresponding to an integrated luminosity of 114~pb$^{-1}$  recently collected by the H1 detector at HERA. 
In absence of a signal, the measurement is interpreted within a minimal model 
parameterised in terms of couplings and compositness scale. New parameter regions, beyond other colliders sensitivities, are explored by the present preliminary analysis.}
\section{Introduction}
The fermion mass hierarchy is one of the greatest puzzles of the Standard Model (SM). It can naturally be explained if the SM fermions are composite, in which case excited states may exist and be produced at colliders.  
A minimal extension\cite{hg} of the SM is used to incorporate excited fermions ($F^*$).
Considering only the electroweak interactions, the excitation part  of the lagrangian is:
\begin{equation}
L_{F^{*}F} = \frac{1}{2\Lambda}{\overline{F^{*}_{R}}}{{\sigma}^{\mu\nu}}[gf\frac{\overrightarrow{\tau}}{2}{\partial}_{\mu}{\overrightarrow{W_{\nu}}}+g'f'\frac{Y}{2}{\partial}_{\mu}B_{\nu}]{F_{L}} + {\mathrm h. c.} ,
\label{lg}
\end{equation}
where the new weights $f$ and $f'$ multiply the SM coupling constants $g$ and $g'$ corresponding to the weak SU(2) and electromagnetic U(1) sectors respectively. The corresponding gauge boson fields are denoted by $W$ and $B$. The matrix $\sigma_{\mu \nu} = (i/2) \left[ \gamma^{\mu}, \gamma^{\nu} \right]$, $\tau$ are the Pauli matrices, and $Y$ is the weak hypercharge.  The compositness scale $\Lambda$ reflects the range of the new confinement force and together with the couplings $f$ anf $f'$ determines the production cross section and the branching ratios of the excited fermions. Effects related to compositness can also appear 
via contact interactions, an alternative not considered here.
\par
  Excited neutrinos can be produced in electron--proton collisions at HERA via the $t$--channel charged current (CC) reaction $e^\pm p\rightarrow \nu^* X$. The cross section is much larger in  $e^-p$ collisions than in $e^+p$ collisions due to the helicity enhancement, specific to CC-like processes.  
  The present analysis uses a data sample corresponding to an integrated luminosity of $114$~pb$^{-1}$ data sample, almost an order of magnitude larger than the previously published analyses at HERA~\cite{Adloff:2001me}.

\section{Data analysis and results}
The excited neutrinos are searched for in the following decay channels: $\nu^* \rightarrow \nu \gamma, \nu Z , e W$. The W and Z bosons are reconstructed in the hadronic channel. The analysis covers 80\% (70\%)  of the total branching ratio for $f=-f'$ ($f=f'$). The selection criteria are described in the following.
\par \underline{\boldmath $\nu^* \rightarrow \nu \gamma $}
Candidate events are selected by requiring missing transverse momentum $P_T^{miss}>15$~GeV. The photons are identified as isolated electromagnetic (e.m.) deposits in the 
calorimeter, measured in the polar angular range $5^\circ<\theta_{\gamma}<120^\circ$. 
The photon candidates measured within the acceptance of the central tracker ($\theta_{\gamma}>20^\circ$) are required to have no associated tracks. 
The neutral current (NC) and charged current (CC) backgrounds are  reduced by imposing the longitudinal momentum  balance $E-P_Z>45$~GeV for events with photon candidates at lower transverse momentum $P_T^{\gamma}<40$~GeV and by requiring the virtuality ($Q_\gamma^2$) computed using the e.m. cluster kinematics to satisfy $\log(Q_\gamma^2)>3.5$. A hadronic jet with $P_T^{\mathrm jet}>5$~GeV is further required in each event. 
\par \underline{\boldmath$\nu^* \rightarrow e W $}
In events with an energetic electron ($P_T^e>10$~GeV) reconstructed in the polar angular range  $5^\circ<\theta_{e}<90^\circ$, the hadronic W decays are searched for by requiring two jets with high transverse momenta $P_T^\mathrm{j1(j2)}>20(15)$~GeV reconstructed  within $5^\circ<\theta_\mathrm{j1,j2}<130^\circ$. The dijet mass should exceed $30$~GeV and the polar angle of the resulting $W$ candidate should be below $80^\circ$.  The background from the NC processes is reduced by requiring the virtuality computed from the electron kinematics $Q^2>2500$~GeV$^2$ if $P_T^e<25$~GeV and by requiring a third jet with $P_T>5$ GeV to be reconstructed in the event if $P_T^e<65$~GeV. 
\par  \underline{\boldmath${\nu^* \rightarrow \nu Z}$}
Candidate events are selected with $P_T^{miss}>20$~GeV and containing at least  two jets with $P_T^\mathrm{j1(j2)}>20(15)$~GeV reconstructed in the polar angular range  $10^\circ(5^\circ)<\theta_\mathrm{j1,j2}<130^\circ$.  The dijet, corresponding to the $Z$ hadronic decay, is required to have an invariant mass  above $60$~GeV.  
 In order to reduce the CC background, the total hadronic system is required to have the polar angle above $20^\circ$ and to contain a  third jet with $P_T^{\mathrm j3}>5$~GeV. The longitudinal balance of the event $E-Pz>25$~GeV is required for events with $P_T^{miss}<50$~GeV. In addition, the topological variable $V_{ap}/V_{p}$ is employed, defined as the ratio of the anti--parallel to parallel projections of all energy deposits in the calorimeter with respect to the direction of the transverse momentum measured with the calorimeter~\cite{Adloff:2000qj}. Due to the multi-jet topology of the signal, large $V_{ap}/V_{p}$ values are expected for $\nu*$ events, in contrast to the CC processes.  Events with $P_T^{miss}<30$~GeV are accepted only if $V_{ap}/V_{p}>0.1$.
\par
The results 
are summarized in table~\ref{tab:nustaryields}. Good overall agreement is observed between data and SM prediction.
\begin{table}[ph]
 \tbl{Observed and predicted event yields for the 
 three event classes.}
{
\begin{tabular}{@{}crr|rrr@{}}
\multicolumn{6}{l}{H1 Preliminary 114 pb$^{-1}$ (e$^-$p Data 2004/2005)}\\
\hline
{} &{} &{} &{} &{}\\[-1.5ex]
Selection & Data & SM & CC-DIS & NC-DIS & ${\gamma}p$ \\[1ex]
\hline                                     
${\nu}^{*} {\rightarrow} {\nu}{\gamma}$ & $12$ & $11.6~{\pm}~2.5$ & $9.1~{\pm}~2.4$ & $1.3~{\pm}~0.3$ & $0.4~{\pm}~0.15$  \\ [1ex]
${\nu}^{*} {\rightarrow} {e}{W_{{\hookrightarrow}qq}}$ & $136$ & $118~{\pm}~22$ & --- & $112~{\pm}~21$ & $4.4~{\pm}~1.2$ \\[1ex]
${\nu}^{*} {\rightarrow} {\nu}{Z_{{\hookrightarrow}qq}}$ & $88$ & $81~{\pm}~15$ & $54~{\pm}~13$ & $5~{\pm}~1.6$ & $22~{\pm}~5$ \\[1ex]
\hline
\end{tabular} \label{tab:nustaryields} }
\vspace*{-13pt}
\end{table}
 \begin{figure}[ht] 
\centerline{
\epsfxsize=4.4cm\epsfysize=4.1cm\epsfbox{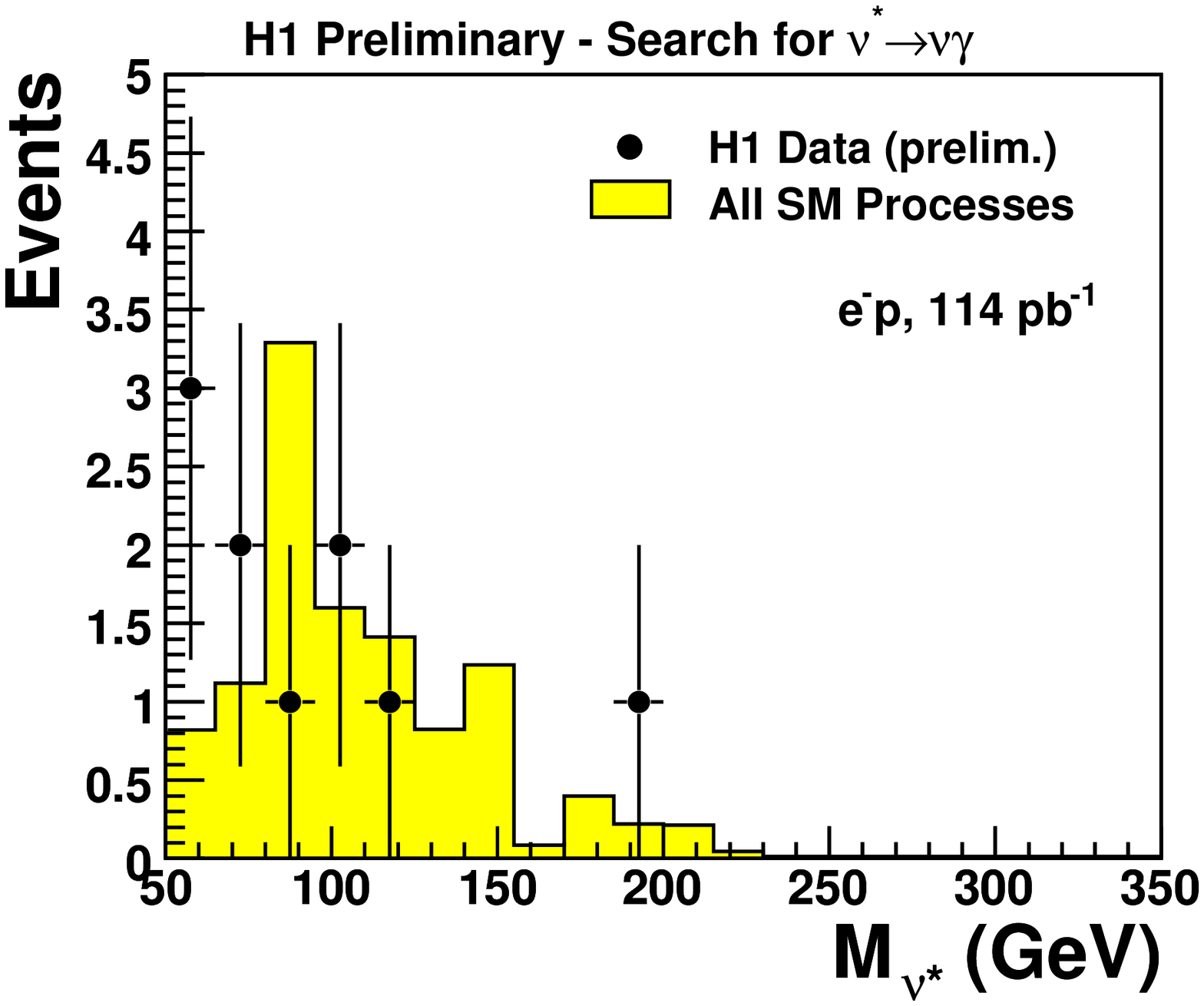}
\epsfxsize=4.4cm\epsfysize=4.1cm\epsfbox{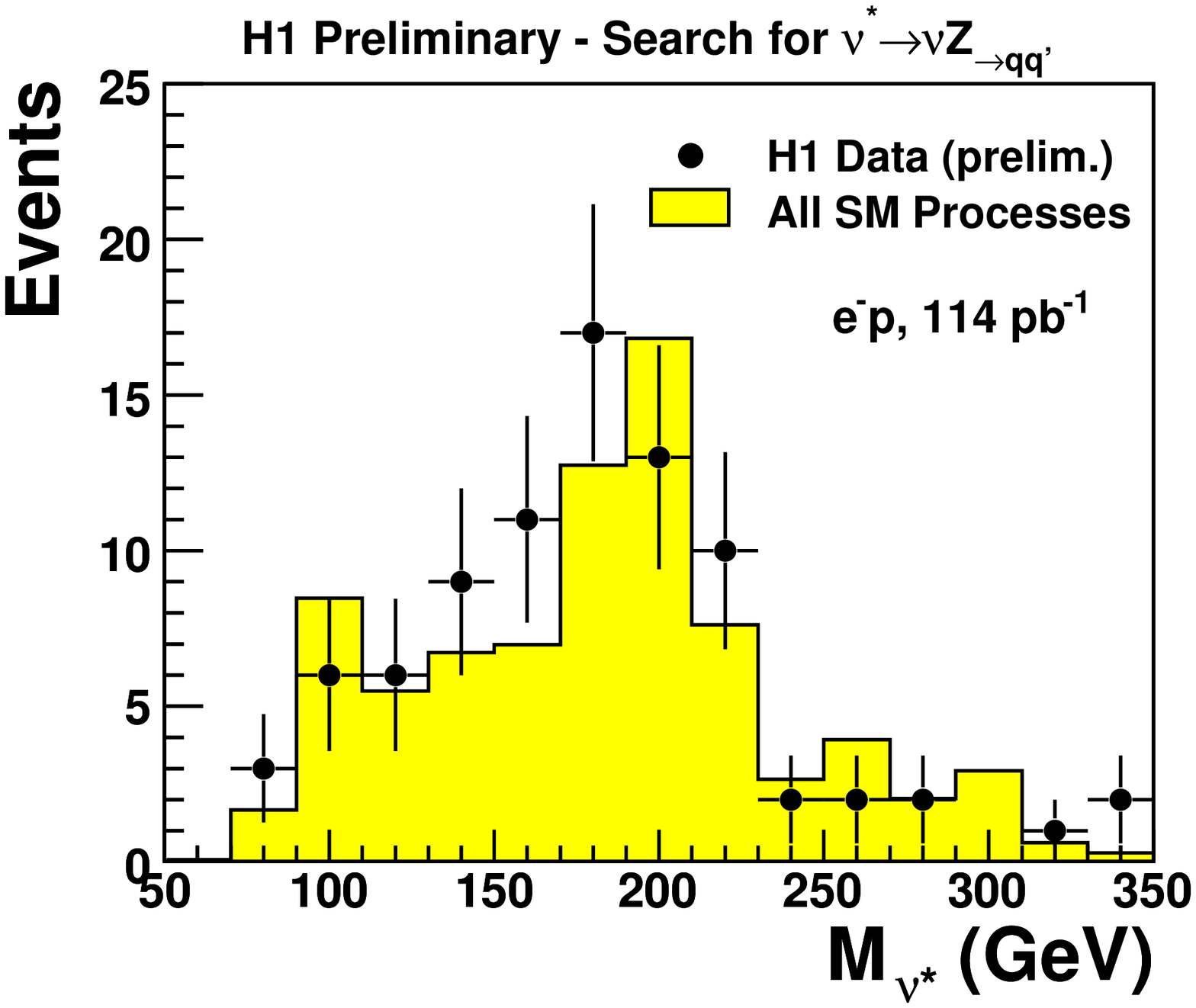}
\epsfxsize=4.4cm\epsfysize=4.1cm\epsfbox{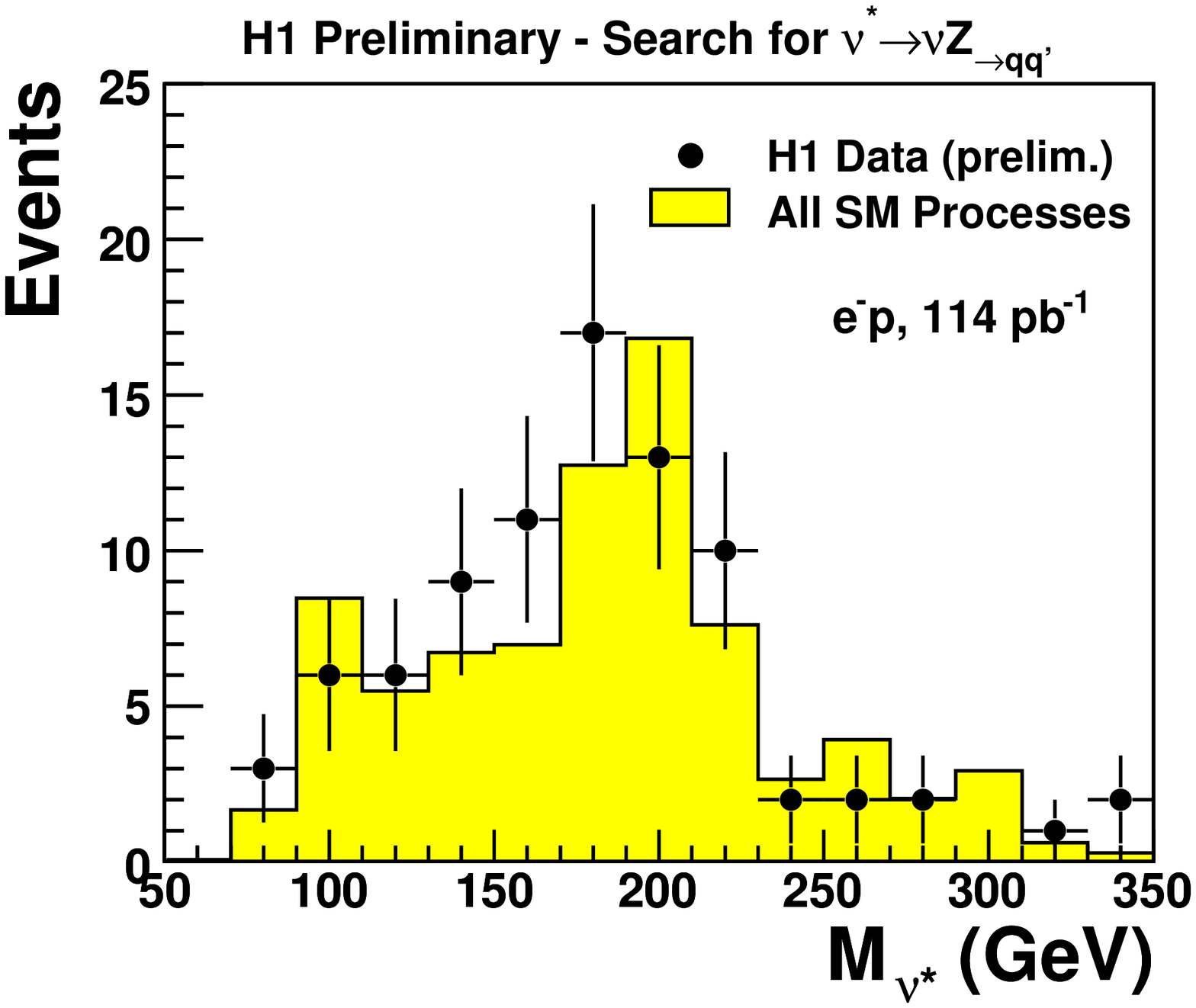}
}
\caption{The invariant mass of the excited neutrino candidates reconstructed in the three decay channels. \label{exnumass}}
\end{figure}
\begin{figure}[ht]
\centerline{
\epsfxsize=5.8cm\epsfysize=5.0cm\epsfbox{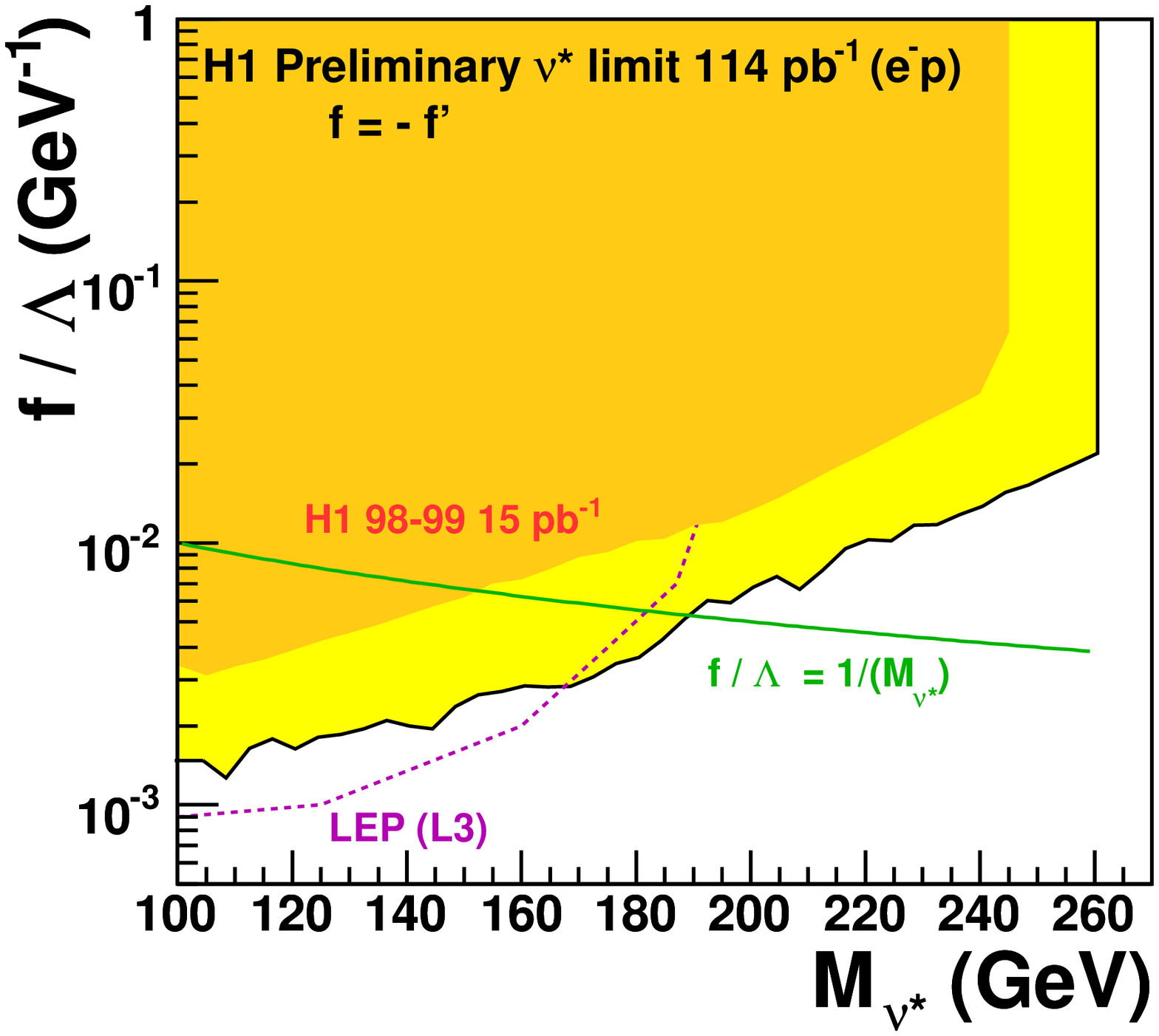}
\epsfxsize=5.8cm\epsfysize=5.0cm\epsfbox{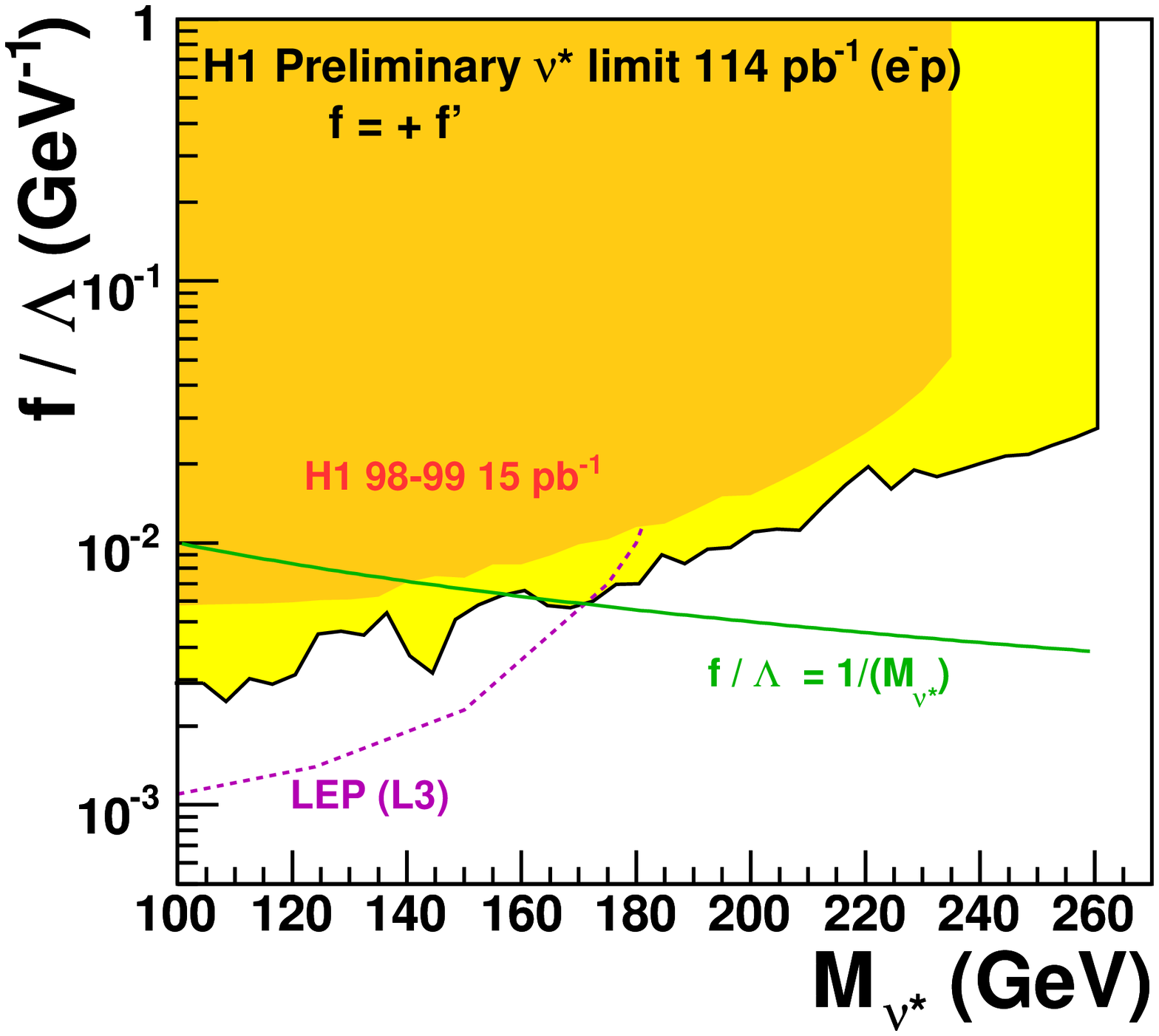}
}
\caption{The limits obtained for the ratio $f/\Lambda$ as a function of the excited neutrino mass within two assumptions: $f=-f'$(left) and $f=f'$ (right, corresponding to a vanishing coupling to the photon and no influence of the $\nu^*\rightarrow \nu \gamma$ channel). \label{fol}}
\end{figure}
For the events selected in the $\nu\gamma$ and $\nu Z$ channels, the neutrino is assumed to be the only non-detected particle in the event and its kinematics is reconstructed assuming the balance of the transversem momenta and the conservation $\sum(E-P_z) =2E^\mathrm{e}_\mathrm{beam}=55.2$~GeV. The invariant mass of the excited neutrino candidates reconstructed in the three channels described above is shown in figure~\ref{exnumass}. No deviation with respect to the SM prediction is observed in these spectra.
In the $M_{\nu *}$ interval [100,260]~GeV
the selection efficiency is about 50\% in the  $\nu\gamma$ channel and varies between
20 and 45\% in the $\nu Z$ and $eW$ channels.
\section{Interpretation and conclusions}
In the absence of a signal for excited neutrino production, limits on the production cross section are calculated using a frequentist approach~\cite{cl}.
The data events are counted in a mass window around a given $M_{\nu*}$ hypothesis and used together with the corresponding SM prediction to calculate an upper limit at 95\% CL on the number of $\nu*$ events, which is then translated into a limit on $\nu*$ production cross section. 
The width of the mass window is varied as a function of $M_{\nu*}$  in order to optimise the expected limit, obtained by replacing the observed by the expected number of events.
The obtained limits on the cross section are translated into exclusion limits in the plane ($f/\Lambda$, $M_{\nu*}$), assuming $f=f'$ or $f=-f'$ (figure~\ref{fol}). For $f=-f'$ (maximal $\gamma\nu\nu*$ coupling) and assuming $f/\Lambda=1/M_{\nu*}$, excited neutrinos with masses below 188~GeV are excluded at 95\%~CL.
\par
The present results greatly extend previous searched domains at HERA and confirm the HERA unique sensitivity for excited  neutrinos with  masses beyond LEP reach.

\end{document}